\documentclass[10pt,twoside]{article}
\usepackage{graphicx}
\usepackage{psfig}
\makeindex

\begin{document}

\thispagestyle{empty}

%\begin{center}
\title{ Gravitational wave experiments and Baksan project ``OGRAN'' }

\markboth{V.N.\,Rudenko et. al.}{Gravitational wave experiments and Baksan project ``OGRAN''}

\author{Bezrukov L.$^{1}$, \\ Popov S.$^2$, Rudenko V.$^2$, Serdobolskii
A.$^2$, \\ Skvortsov M.$^3$
\\[5mm]
\it $^1$Institute for Nuclear Research, \\ \it Russian Academy of Sciences, Moscow, Russia\\
\it e-mail: bezrukov@inr.ac.ru\\
\it $^2$Sternberg Astronomical Institute, \\ \it Moscow State University, Moscow, Russia\\
\it e-mail: serg@sai.msu.ru , rvn@sai.msu.ru\\
\it $^3$Institute of Laser Physics, \\ \it Siberian Branch of Russian Academy of Sciences,
\\ \it Novosibirsk, Russia\\
\it e-mail: skv@laser.nsc.ru
}

\date{}
\maketitle

%%\end{center}

\begin{abstract}
\noindent
A brief sketch of the present status of gravitational wave experiments is 
given. Attention is concentrated to recent observations with the gravitational
detector network. The project OGRAN for a combined optic-interferometrical
and acoustical gravitation wave antenna planned for installation into
underground facilities of the Baksan Neutrino Observatory is presented. We
describe general principles of the apparatus, expected sensitivity and current 
characteristics of the antenna prototype; some ways for sensitivity
improvement are also discussed.
\noindent

{\bf Keywords: gravitational waves, gravitational detectors.}
\end{abstract}

\section{Introduction}

Current status of the gravitational wave (GW) experiments is characterised
by the well developed international net of gravitational wave antennae of two
types: optical interferometers on suspended mass-mirrors and cryogenic
acoustical resonant bars with the SQUID sensor. ``Bar net" containing five
independent detectors presented by IGCE \cite{ref1} and ROG \cite{ref2}
collaborations is well developed and has a relatively long history of
observation. ``Interferometer family" composed by LIGO \cite{ref3}, VIRGO
\cite{ref4}, GEO \cite{ref5} and TAMA \cite{ref6} recently started first
individual and joint series of observations on some intermediate sensitivity
but a work for improving it is going on. 

The modern cryogenic resonance bar detectors now are able to register
metric perturbations at the level $10^{-21}$ in a narrow bandwidth on the
order of few Hertz around the resonance frequency. Large base 
gravitational-wave interferometers currently can detect GW-bursts with the magnitude
$\sim 10^{-19}$ and spectrum width of several hundred Hertz. It is believed
that in a nearest future the free-mass interferometrical antennae will achieve
the wideband sensitivity $10^{-22}$ which corresponds
to the expected rate of events at least one per day and more. Approximately
similar sensitivity might be reached by ``big ball" detectors having the mass
$30\div 60$ ton cooled up to 10 mK \cite{ref7} with advantage of the increasing
expected rate of events due to the omni-directional antenna pattern.
At present this type of detectors exists as a moderate scale
prototypes \cite{ref7,ref8,ref9} and a funding of the full 
scale projects is not solved completely. Thus one could say there is some
``scientific competition" between solid body resonance acoustical detectors
at one hand and wide band free-mass optical interferometers on the other
hand. Despite of a visible potential advantage of the interferometers
(excepting their cost which is ten times more the cost of bar detectors)
both types in fact have a supplementary relation to each other taking into
account the different mechanism of their interaction with gravitational
radiation: a generation of acoustical waves for the bar and excitation of
mass-mirror's relative displacements for interferometers (in the last case
the effect can be described also as variations of the optical wave length). 
This difference and also the observation some anomalous coincident events 
with bar detectors \cite{ref10,ref11} stimulated a discussion for 
a comparison of the cross sections both type of gravitational antennae
which had been started after supernova SN 1987A event \cite{ref12,
ref13} and renewed recently \cite{ref14} after results published 
by ROG collaboration. 

Meanwhile joint observations with bars and interferometers at the current 
level of sensitivity were carried out in past and its are continuing 
at present \cite{ref15}.

It was mentioned in many papers, see for example \cite{ref16}, that a  
sensitivity threshold in the GW searching is much higher the potential
sensitivity of any individual detector due to the ``blind character of 
all sky searching" when coordinates and parameters of a hypothetical source 
are unknown and one has to run over all conceivable positions and signal 
templates in a data processing. It increases a ``chance probability" to get
a false ``noise event" or equivalently decreases the effective registering
sensitivity. Although the declared aim of the present antenna net is an
operation at the level of $10^{-21}$ the typical effective threshold of
``bursts sensitivity" resulted in many recent observations is still occurred 
between $10^{-18}$ and $10^{-19}$ magnitude of metric perturbation. 
Such moderate sensitivity allows to overlap mainly sources located in a
close environment of the Galaxy with the radius less then $1\,Mpc$ i.e. 
this type of GW experiment can be called as a ``search for rare events" with
a hope to have a ``lucky chance" of nearby ``relativistic star catastrophe".

In the papers \cite{ref17,ref18} it was proposed that the similar
sensitivity might be achieved for a room temperature bar detector with an 
optical read out using an advanced optical technology (low noise optical
frequency standards). Group of the cryogenic antenna AURIGA has  
developed a micro-gap optical transducer as a low noise alternative to 
the SQUID read out \cite{ref19}. The authors of the paper \cite{ref17} have
analysed a more sophisticated variant of the optical read out with the 
optical FP-resonator extended along the bar body and in fact having the
same scale as the bar itself. One can consider this construction as a 
combination of the bar and interferometer unified in the one detector 
and wait a more complex response under GW excitation containing the
acoustical and optical parts independently in some extent. This idea
now is in the process of realization in the form of Russian national
project OGRAN (opto-acoustical gravitational antenna) \cite{ref20}.

In our report below we at first, review recent results collected by the
existing net of GW antennae and at second, present more in details a current
status of the OGRAN project.

\section{Results of the GW searching}

The principal filtration procedure in a searching for GW pulses
(introduced by a pioneer of the gravitational wave experiment
J.~Weber \cite{ref21}) consisted in a selection of coincident perturbations
of widely spatially separated detectors. It is the straightforward method for
an effective decreasing of the local noise background but more of that the
coincidences (as it was believed) could guarantee
in some extent a global (extraterrestrial) nature of registered signals.
This procedure remains to be valid now but it is enriched (if not loaded)
through a preliminary reduction of the detection noise by optimal filtering
matched with a definite type of expected signal (see a short 
review in \cite{ref22}).

The following ``signal families" are considered more often: a) relativistic
binary coalescence, b) supernova explosions (or more general,-collapses),
c) pulsars, d) primordial stochastic GW background. However under a deficit
of ``a prior information" (no data of source's position and parameters) the
algorithm of matched filtering for such ``blind search" becomes extremely
cumbersome, requires a lot of computational time and power etc. It affects
an ability of quick online detection and finally results in a decreasing of
effective sensitivity. Nevertheless just in this manner a number of recent
observations were performed. Results of some of them are presented below.

{\bf {i) Interferometers.}}

LIGO collaboration has published a series of articles after the first
short observational period called as ``the first Science run, S1": (23.08 -
09.09) 2002. Three interferometrical detectors were involved: two in the
Hanford Observatory with the $2\,km$ base (H1) and $4\,km$ (H2) \cite{ref22}
and one $3000\,km$ apart in the Livingstone Observatory also with the
$4\,km$ long base \cite{ref23}.

a) Result of the {\it{search for short GW bursts}} \cite{ref24}.

This program is more close to the original Weber's algorithm of selection of
very short coincident pulse-excitations at outputs of independent detectors.
A typical new element introduced at present consists in a modelling of
expected signals in two probable waveforms: a quasi $\delta$-pulse with
Gaussian envelope, and similar pulse with a sine carrier,-the ``sine-Gaussian"
pulse. The carrier frequency and pulse time width (duration) were varied.
So the ``matched filtering" could be applied only as a multi-channel procedure.
A preliminary noise filtering and selection of ``GW-burst candidates" also was
very sophisticated. It is reflected as a step by step reduction of the
useful observational time interval \cite{ref24}: 17 days (408 hours) (S1 time)-
96 hours (effective duty cycle) - 35.5 hours (triple-coincidence time).

The final conclusion was in the determination of the ``first upper limit
from LIGO on GW bursts". Roughly it was formulated as a statement that
``with the confidential level $90\%$ at the frequencies $(150\div 3000)\,Hz$
short GW bursts with duration $(40 \div 100)\,ms$ and spectrum amplitudes
$h_{f}\sim (10^{-18}\div 10^{-19})\,Hz^{-1/2}$ can be expected with the rate
$ R\leq 1.6\, day^{-1}$". The observation was performed in several frequency
windows $\Delta f \simeq 10^{2}\,Hz$ inside the general bandwidth. Thus
the threshold for admitted metric perturbations has to be defined as
$h\leq 10^{-17}\div 10^{-18}$. Below we shall compare it with bar's
observations.

b) Search for {\it{GW from binary neutron stars coalescence}} \cite{ref25}.

This program is known as a detection of ``chirp signals", i.e.
short GW pulses with sweeping carrier which have to be radiated by
a relativistic binary during of the last orbiting cycles of coalescence.
A standard example of coalescence for a neutron star binary
$2 \times 1.4\,M_{\cdot}$ provides the chirp signal which would traverse the
sensitive band of the interferometers $(100\,-3000)\,Hz$ in 2 seconds.
At the current sensitivity the Livingstone detector was able to register
such signals from the distance $\sim 176 Kps$ and the Hanford instrument
from $46\,Kps$. So the only the Milky Way events might be overlapped
in this experiment with a small contribution $\sim 10\%$ from the Large
and Small Magellanic Clouds.

The bank of templates used in this observation was modelled for variable
parameters such as: $i$-the orbit inclination, $\alpha$-the initial
signal phase and masses $(1\div 3)M_{\cdot}$. A ``density" of templates was
chosen so that a loss of SNR did not exceed $3\%$; it gave a total number
of templates close to 2100. Noise characteristics were studied on the base
of arbitrary selected lumps of noise playground; special sophisticated
veto criteria were applied to avoid a statistical bias.

During the observational S1 run no candidates for GW-chirps were found
at the threshold SNR=8 corresponded to the $90\%$ of confidential limit.
Meanwhile a Monte-Carlo simulation with an inserting of artificial chirp
-signals confirmed an ability of detection of coincident events. So a
final conclusion was formulated as "a new upper limit of the neutron star
binaries coalescence rate for Milky Way Equivalent galaxies is $R\leq
1.7\cdot
10^{2}\,year^{-1}$" which is better then previous results but is still far from 
a realistic astrophysical prediction.

The designed sensitivity of LIGO set-ups would allow to overlap binaries
at the distance $\sim 21\,Mpc$. Under the optimistic theoretical vision
of the galactic rate of coalescence $5\cdot10^{-4}\,year^{-1}$ \cite{ref26} it
forecasts the integral rate of chirp-events $0.25\,year^{-1}$.

c)Search for {\it{a continuous GW radiation (pulsars)}} \cite{ref27}.

There are several theoretical scenarios forecasting the pulsar's
gravitational wave emitting. It may be an NS-spin precession, oscillations of
eigen modes, r-modes crust currents ect. but as it's believed the more
effective GW radiation would be produced by its non-zero ellipticity
$\varepsilon \,< \,1$, which is defined by a relative asymmetry of 
inertia momentum in the plane normal to NS rotational axis. Available
in literature theoretical estimates of the ellipticity predict as a 
realistic value $\varepsilon \sim 10^{-7}\div 10^{-8}$ although  for NS with
a very strong magnetic field it might be several orders high \cite{ref28}.
At any case the astrophysical expectation GW radiation from galactic pulsars
does not exceed $h \sim 10^{-24}$ metric perturbation.

During the S1 run LIGO collaboration carried out a test of GW observation of
the millisecond pulsar $J1939+2134$. This type of measurement was associated
with long time accumulation of the narrow band $\Delta f \simeq 4\,Hz$ 
antenna's output noise centred at the double pulsar frequency
$2f_{ps}\simeq 1284\,Hz$. Before accumulation the raw output data were 
demodulated through a special time resampling and after its were
undertaken to a very complex multichannel ``matched filtering" procedure with
an appropriate bank of templates varied through unknown parameters of
the source (orbit inclination, polarisation angle and initial phase). 
There was no a "coincidence procedure" in this measurement of the continuos
signal from well positioned source, results of all detectors were used
independently and might be summed (also data of GEO set up were involved).

The lowest strain noise spectrum density of the L and H detectors in these 
measurements were at the level  $\sim 10^{-22}$. Effective time duty cycle 
was $209\,h$ for the Hanford and $137\,h$ for Livingstone instruments.
The upper limit for the detected amplitude of GW radiation of the pulsar 
$J1939+2134$ inferred from S1 run data consisted in $h\simeq (1.5\div 4.5\,)
10^{-22}$ (accounting for different interferometers involved). This limit
does not present any new qualitative knowledge for astrophysics but it was
a first time when the measurement was performed with the well defined target,
- the concrete pulsar with known coordinates and frequency. Estimation of its
ellipticity resulted from this upper limit also was large $\varepsilon\,< \,3
\,10^{-4}$ and would require a huge magnetic field $\sim 10^{16}\,G$.

d) Detection of the {\it{GW stochastic background}},
\cite{ref28,refA3}.

This is a very ambitious program of detection a relic GW stochastic
radiation produced in a much more early universe (at the ``Planck time")
then its electromagnetic analog CMBR.
Such objective looks more problematic in respect of the current
interferometer sensitivity comparing with the above mentioned programs.
A conventional description of that backgroung contains a special parameter 
$\Omega_{g}(f)$ which is the GW energy density per unit logarithmic frequency 
normalized on the critical energy of the ``close universe" $\Omega_{g}(f)=
(f/\rho_{c})(d\rho_{g}/df)$. This parameter is related with the power
spectrum of the gravitational wave strain $<h^{2}(f)>$ via the 
formula \cite{ref29}: $ <h^{2}(f)>=(3H^{2}/10\pi^{2}\,f^{3})\Omega_{g}(f)$,
where $H\sim (1.5\cdot 10^{10}y)^{-1}$ is a present day Habble constant.
Thus the largest $h(f)$ corresponds to the low frequency wing of the
interferometer bandwidth, so for $f=100\,Hz$ one comes to the estimate
$h(f)\sim 5\cdot 10^{-22} \sqrt{\Omega}_{g}\,Hz^{-1/2}$ with $\Omega_{g}
\simeq Const.$ In the Standard Cosmological Model the GW relic background
is considered as an isotropic, stationary, Gaussian noise with $\Omega_{g}
\leq 10^{-8}$ \cite{refA3}.

A detection method consists in a measurement of the cross-correlation
variable combined from outputs of two independent detectors located in
the same place. Then a correlated ``GW noise signal" has to be accumulated
in time at the background of independent intrinsic noises of both detectors
The sensitivity to the parameter $\Omega$ growths as $\Omega \sim |h|^{2}
/\sqrt{\Delta f\, T}$ where $T$ is the time of measurement (integration time).
This ideal theoretical scheme needs a correction associated with real
positions of GW-interferometers separated by a known distance. The correction
factor $\gamma$ decreasing the sensitivity in the case of H and L
interferometers was $\gamma \sim 0.1$ at the frequency $\sim 100\,Hz$.

During the S1 run only $\sim 100\,hrs$ of coincident interferometric
strain data were selected to establish an "LIGO upper limit" on stochastic
background of gravitational radiation. With the $90\%$ confidence limit
at the frequency band $100-300\,Hz$ the experimental estimation of
$\Omega_{g} h_{100}^{2}\le 23$, where $h_{100}$ is the Habble constant in
units of $100\,km/sec/Mpc$. This new limit is not too useful for a testing
astrophysical models but it is much better then the previous result from
interferometers \cite{ref29}.

{\bf {ii) Bars.}}

Bar detector collaborations IGEC (International world net) and ROG (Italian 
net) also published reports in respect of the programs mentioned above.
In compare with interferometers the ``bar observations" have an advantage of
much more long duty cycles but a lack of narrow bandwidth. Upper limits
for different type of GW signals collected with bar detectors briefly
are summarised below.

In the program of {short GW bursts search} data of five cryogenic bars
were analysed at the four year time interval $1997-2000$ (there were
three Italian set ups, one of US and one Australian) \cite{ref1}. 
A distribution of the effective observation time was the following: during
4 years there were 1319 days when at least one detector was operating;
707 days with at least 2 detectors in simultaneous operation; 173 days with
3 detectors and 26 days with 4 detectors (5 detectors practically
had no the overlaping time of joint operation). In these experiments a strength
of the expected GW bursts was quantified through its
Fourier amplitude $h_{f}\simeq(1/4Lf^{2})\sqrt{E/M} \, Hz^{-1}$ defined
by $E$ - the energy deposited by a GW burst in the detector with parameters:
$M$-mass, $L$-length, $f$- mean resonance frequency. Four detectors had
frequencies close to $900\,Hz$ but Australian detector NIOBE had $f\sim 
700\,Hz$. The matching filtering was performed referring to the $\delta$-pulse
modelling of GW bursts.

The result of the blind search declared by IGEC sounds as ``no coincidences
have been found below the threshold $h_{f}\simeq 10^{-20}\,Hz^{-1}$;
at the threshold itself the registered ``noise event" rate was ``at least one
per year". It means a daily rate ot GW bursts above this threshold has to be
less than $\sim 4\times 10^{-4}$, that looks much stronger the LIGO result.
However to get an estimation of the integral magnitude of metric
perturbation (used in the LIGO approach) one has to multiply Fourier
amplitude to the burst's spectrum width. For "short bursts" it is the
number comparable with bar's resonance frequency $\sim (10^{2}-10^{3}\,Hz$.
Also the event rate has to be corrected on the factor of ``ratio of
bandwidths" $\sim (2\,\div\,3) 10^{3}$. Then the reduced ``IGEC upper limit":
``bursts with amplitude $h\leq 10^{-18}\div10^{-17}$ might have the rate
$R\leq 1\,day^{-1}$" and becames comparable with the LIGO result mentioned
above. At the same time a ``robustness of IGEC limit" is much high the
same of LIGO because the ``multichannel (triple etc.) coincidences" 
exponentially decrease the ``false alarm" error.

For bar detectors there is no a special (separate) program type of {\it{a
search GW signals from binary coalescences}}. Due to a very narrow frequency
bandwidth of these instruments such program naturally is unified with the
``coincident bursts search".

In the program of {\it{searching for periodic GW signals}} ROG collaboration
has got the more stronger ``experimental upper limit"  then LIGO.
In the paper \cite{ref30} 2 day blind-sky search was performed with the
EXPLORER's data of 1991 year. The search was centred at the frequency
$922\,Hz$ in the band $\sim 1\,Hz$. It resulted in the upper limit
on the order of $\sim 3 \times 10^{-23}$ with $99\%$ confidential level.
The same data were used in the paper \cite{ref31} for searching periodic
signals from the Galactic center in the bandwidth $0.06\,Hz$. For 95 days
observation the upper limit of the ``incoherent searching" (i.e. without
any template) was estimated as $\sim 3\,10^{-24}$ \cite{ref31}.
A similar searching targeted to Galactic center was performed by the ALLEGRO
detector group \cite{ref32} at frequencies $896.5,\,920\,Hz$ in the band
$\sim 1\,Hz$. The upper limit $h\leq 8\cdot 10^{-24}$ was found.

These results are roughly one order of magnitude stronger the LIGO one.
But the principal difference is the LIGO observation was well targeted
coherent search with the definite source and family of templates covering
a bank of unknown spin-down parameters. Such search physically is more
informative. It allowed in particular to estimate bounds for pulsar's
ellipticity (see above).

At last a limitation of {\it{GW stochastic background}} was also derived 
from observations at the bar couple EXPLORER/NAUTILUS \cite{refA3,
ref3A}: it was found $\Omega_{g}\leq 6\cdot 10$ at the very narrow 
frequency interval $907.15\div 907.25\,Hz$; where the factor ``10" is the 
result of observation but the factor ``6" takes into account the distance 
between the bars $\sim 600\,km$ (i.e. it's proportional to value 
$\gamma^{-1}$).

Thus one can conclude the all upper limits established in recent experiments 
with gravitational wave antennae are still rough for a checking of validity 
astrophysical models used to forecast the power and rate of expected GW events. 
The effective threshold of metric variations derived from these experiments
was not below the level $h \sim 10^{-19}$ for the ``short burst" detection
and $h \sim 10^{-22}$ for continuous signals. One can wait an essential 
improvement of these numbers from advanced interferometers in a nearest
future. Meanwhile a special strategy of observation so called a ``search
for astro-gravity correlation" makes it reasonable to continue experiments
with available antennae even at its current sensitivity level.

\section{Search for GW events associated with GRB}

This approach is one of the more developed particular version of the common
program of ``searching for GW events accompanied by other types of radiation"
which can be registered by parallel observational channels such as neutrino
telescops, x-ray and cosmic particles detectors \cite{ref33}. The idea of
such strategy is based on the general principle of optimal filtration of
weak signals: the more {\it{a prior information}} is accessible the better
a quality of detection. A modern understanding of the nature of two
astrophysical phenomena, gravitational wave events and Gamma-Ray Bursts
(GRB), suggests that both phenomena may have common progenitors - superdense
relativistic stars at the moment of some catastrophic processes in their
evolution just as binary coalescence, stellar core collapse, fragmentation
etc. (see for a GW review \cite{ref34}, for GRB models \cite{ref35},
and references therein). Thus it is reasonable to look for anomalous
in the output noise of gravitational detectors around time marks defined
by the registered GRB. It strongly reduces the amount of stochastic
data under processing compare with the ``blind search" strategy and besides
provides a possibility to integrate weak GW signals through an optimal
summing over many GRB events. Then one may hope to discover GW-bursts even
with antennae of the moderate level of sensitivity.

The problem of such parallel multichannel searching firstly was considered
in \cite{ref36} for bar detectors and in \cite{ref37} for interferometers.
In the papers \cite{ref38,ref39} a nontrivial problem of accumulation
GW-GRB signals was analysed. The effectiveness of the accumulation directly
depends on a ``prior information" of the GW burst's time position with respect
to the GRB. Just in this point there is a large uncertainty of theoretical 
models.

Nevertheless some algorithms of the ``cumulative detection" have been
proposed accounting for the GRB time structure \cite{ref40} and the signal 
processing typical for cryogenic bar detectors.

First experimental tests of checking the hypothesis of GW-GRB association 
are presented in the papers \cite{ref41,ref42}. Still no positive 
results were reported. The current upper limit for GW burst amplitude of 
this type of signal established by the ROG group is  $h\leq 2.5\cdot 10^{-19}$ 
\cite{ref43}.

\section{Search for rare events}

The deficit of sensitivity of the available detectors obliges
experimentalists to be in the frame of program a ``search for rare events",
i.e. to keep antennae in a long duty cycle relaing to a ``lucky chance" of
relativistic catastrophic processes in nearby region $r\leq (50-100)\,
kpc$ of our Galaxy. This situation is similar to modern programs of neutrino 
astrophysics concerning a ``searching for star collapse" by registering
a cosmic neutrino radiaztion. The sensitivity of all known neutrino detectors
allows a $\nu$-flux registration from sources located close then $100\,kpc$.
Nevertheless precedents of such events had took place in last years.

The first one was occured during the SN1987A when some coincident signals 
registered by room temperature bars \cite{ref10} were assosiated with
neutrino bursts detected simultanously by neutrino telescops.
Despite of the very strong criticism \cite{ref44,ref45} this case
had definitely demonstrated a reason of the ``astro-gravity correlation" 
stratedy and much more a reason to keep in continious operation detectors 
even of modetare sensitivity.

The second case was reported recently by ROG collaboration. Analysis 
{\it{a posteriori}} of
the EXPLORER-NAUTILUS data collected in the year 2001 discovered an excess
of statistically meaningfull coincident exitations \cite{ref46}.
The excess observed in the sidereal time coordinate was concentrated around
the four hour time mark with width two hours. At this period the two bars
were oriented perpendiculary to the galactic plane and their sensitivity
for galactic GW-sources was maximal. The energies of coincident events
deposited in each bar were aproximately equal and being recalculated to
the radiated energy were estimated as $\sim 10^{-2}\,M_{\odot}c^{2}$ for
sources in the Galaxy center. At the same time the rate of events was
too high $\sim 200/y$ in compare with conventional astrophysical
forecasts. Afterwords the statistical significance of this observation
was critically discussed \cite{ref47}, \cite{ref48} but recent
observations it seems confirm the first result. Detailed theoretical
analysis \cite{ref49} still did not lead to a definite physical explanation
of the phenomenon: it looks as there would be unknown sources distributed in
the galactic disc or a few (if not one) closly located unvisible GW-burst 
repeters. An exotic supposition of GW radiation produced by primordial black
hole binaries contradicts to the requirement of homogenious distribution
of such objects in the Halo. Thus observations have to be continued.

\section{Opto-acoustical gravitational antenna}

It follows from our brief review above the current ``effective upper limits"
on cosmic GW pulses consist $(10^{-19} \div 10^{-18})$ in term of metric
perturbation. It is interesting to note that interferometers achieved
such sensitivity without a deep cooling (only advanced versions are planned
to be with ``cryogenic mirrors" \cite{ref50}). This is the result of the two
technical ideas realized into constructions of these set ups. The first is
the operation at frequencies apart from mechanical resonances of ``high Q
suspensions" where the brownian noise is strongly suppressed. The second
in using the laser optical read out with a very small back action of
the photon short noise: it allows to amplify the opto-mechanical 
transformation of mirror's displacements by encreasing the laser power 
(one can remark here that up to now super cryogenic bars did not realize 
its potential sensitivity limited by noises of the SQUID read out 
\cite{ref51}).

This understanding was put in a basement of the Russian national project
OGRAN having foreseen the construction of a room temperature bar combined
with FP-interferometer as a readout \cite{ref20}.

%\begin{figure}[h]
%\centerline{\hbox{\includegraphics[width=0.5\textwidth]{rudenko_fig1.eps}}}
%\caption{Principle layout of the OGRAN detector.}
%\label{fig1}
%\end{figure}

The principal scheme of the OGRAN is presented on figure 1.
%\ref{fig1}.
It consists of the bar with a tunnel along central symmetry axis where
a high finess FP interferometer is formed by two mirrors attached to the
bar's ends. In contrast with the optical read out developed for the AURIGA
by the Legnaro group \cite{ref19} such construction has the expanded
optical cavity with bar's length instead of a microgap optical sensor
attached at one end of the bar. This difference has a qualitative
consequency: in general the responce of such combined opto-acoustical
antenna must contain besides acoustical exitation of the bar also an
``optical part" as a result of GW-EM interaction. A theory of this antenna
was considered in \cite{ref17}. It was concluded that in a free-mass
interferometer the ``optical" and ``acoustical" parts were undistiguished,
but for the ``bar-interferometer" with mirrors following nongeodesic paths
their difference has to be observable. One must to recognize the
difference tends to zero in a ``very long GW-approximation" nevertheless this 
new important feature of the combined antenna has to be taken into accout
with an open mind.

A principal possibility to achieve the resolution $h\leq 10^{-18}$ at the
bar with the optical sensor was proved in the paper \cite{ref18}. Briefly 
the argumentation was as follows.

The realistic sensitivity of the bar antenna for GW bursts with duration 
$\tau$ is determined by the general formula
\begin{equation}
h_{min}\ge 2L^{-1}(kT_{n}/M\omega_{\mu}^{2})^{1/2}\cdot(\omega_{\mu}
\tau Q)^{-1/2}
\label{1}
\end{equation}
where $L,M,\omega_{\mu},Q$ are the bar parameters: the length, mass, resonant
frequency and quality factor; $k$ is the Boltzmann constant. The $T_{n}$
is the effective noise temperature which depends upon transducer and
amplifier noises and coupled with the physical temperature $T$ through a
noise factor $F$: $T_{n}=T\cdot F$. The substitution in {\label{1}} of the
typical parameter's values: $L=2\,m\,$,$\,M=10^{3} kg\,$,$\,Q=3\cdot10^{5}$,
$\omega_{\mu}\simeq 10^{4} s^{-1},\,\tau=10^{-3}s,\,T=300\,K$ resulits in
\begin{equation}
h_{min}\ge 1.5\cdot 10^{-19}\cdot F^{1/2}
\label{2}
\end{equation}
The noise factor $F$ is defined by the ratio a real noise
variance to the thermal noise variance in the antenna bandwidth $\Delta f
\simeq \tau^{-1}$. For the optical interferometer read out
\begin{equation}
F=(2M/\tau)(G_{e}/G_{b})^{1/2}
\label{3}
\end{equation}
Above the spectral densities of Brownian $G_{b}$ and optical $G_{e}$ noises
were introduced with the definitions  $G_{b}=2kTM\omega_{\mu}/Q$ and
$G_{e}=B \omega_{\mu}^{2}(2\hbar\omega_{e}/\eta W)(\lambda_{e}/2\pi N)^{2}$;
where $\omega_{e},\,\lambda_{e},\,W$ are the frequency, wave length, power
of the optical pump; then $\eta,\,N,\,B$ are the photodiod quantum efficiency,
number of FP reflections and ``excess noise factor" (number of times a real
optical noise exceeds the short noise level). For the designed OGRAN
optical parameters: $W=(1-3)\,Wt, B\simeq (1-10),\,\lambda_{e}=1.064\,\mu,
\theta=0.8,\,N=(10^{3}-10^{4}$ one can find the estimation $F\simeq 1$,
so the forecasted sensitivity (2) reduces to $h\sim 10^{-19}$.

It is important to emphasize that the formula (1) supposes a whitenning
of the Brownian bar noise (a cut off the bar's resonance noise
region) and operation at the ``wings" of the thermal noise spectrum.
A response to GW exitation out of the resonance is very small but the
modern optical read out is capable to pick it up.

A practical realization of this opto-acoustical detector associated with
a high frequency stability laser as a source of the optical pump. A simple
direct injection of the beam into bar's FP resonator would requier the
unrealistic frequency stability according with $ \Delta\omega/\omega=h\sim
\,10^{-19}Hz^{1/2}$. For this reason the practical scheme has to be composed
as a ``differential bridge" with automatic compensation a large part of
frequency drifts. Fom the experimental optics two types of such ``bridges" 
are known: first is the Michelson interferometer (just it was taken for the
interferometric GW antennae); second is called as a ``comparator of optical
standards" in which one narrow frequency EM source refers to the similar
one and slow drift of both might be corrected. This type of scheme was used
by the Legnaro group \cite{ref19} and it was choosed also as a ``preferable 
technique" for the OGRAN set up.

%\begin{figure}[h]
%\centerline{\hbox{\includegraphics[width=0.5\textwidth]{rudenko_fig2.eps}}}
%\caption{LPF-scheme of laser frequency stabilisation:
%%\protect\\
%EOM --- electro-optical phase modulator;
%DBM --- double balance mixer;
%AFC --- laser frequency driver;
%PD --- photodetector.}
%\label{fig2}
%\end{figure}

The OGRAN collaboration consists of three Russian Institutions:
{\it{i)}} Sternberg Astronomical Institute of Moscow State University,
{\it{ii)}} Institute for Nuclear Research RAS,
(Moscow), {\it{iii)}} Institute of Laser Physics Siberian Branch RAS
(Novosibirsk). The declared goal was to install the large ($\sim$ 2.5 ton) 
opto-acoustical GW detector into undeground camera of the Baksan 
Neutrino Observatory (INR) and to use it in a duty cycle in cooperation with 
the GW world net firstly at room temperature and then in its advanced 
cryogenic version. ILP keeping a leading position in Russia as a center of
high frequency stability optical standards dveloped a specific variant of 
the laser frequency stabilisation by a high finess optical cavity 
{{\ref{fig2}} which was selected for the OGRAN set up.

A logical structure of the OGRAN measurement scheme contains a frequency
coupled laser source and FP-cavity of the bar, so that oscillation of the
bar's length resulted in frequency variations of the output light beam which
has to be measured by some discriminator based on a very stable external
optical cavity. Idealogically this correspods to the AURIGA 
optical sensor \cite{ref19} but its realization becames much more complex 
for the long (expanded) cavity then for the micogap one.

%\begin{figure}[h]
%\centerline{\hbox{\includegraphics[width=0.5\textwidth]{rudenko_fig3.eps}}}
%\caption{The principal layout of OGRAN project:
%%\protect\\
%EOM --- electro-optical modulator;
%45P --- polarizer;
%$\lambda /4$ --- plate $\lambda/4$;
%FD --- photodetector;
%Mrs --- adjustable mirrors.}
%\label{fig3}
%\end{figure}

A first prototype of the OGRAN was developed in the SAI MSU. Its optical
scheme is given on figure 3.
%{\ref{fig3}}.
Only small effective laser power
$\sim 2\,mW$ was available, so the two mirrors FP cavities were utilized
with Faradey isolators instead of the ``three mirrors" resonator shown at
the {\ref{fig2}} (for a large laser power the three mirror construction is
preferable). In the mecanical part a small pilot model of the bar detector 
was manufactured with parameters: $M\simeq 50 kg,\,L=50\,cm,\,\omega_{\mu}=
2\pi\,\times\,5 kHz$. The finess of FP-cavities of the bar and discriminator 
were equal $F\simeq 800$. The discriminator was thermoisolated and had a feed 
back loop at slow frequencies to tune a position of one mirror (attached with 
PZT to its end) keeping the optical resonace. Calibration experiments
have shown the absolute sensitivity to the pilot bar oscillations at the 
level
$\sim (1-2)\cdot 10^{-14}\, cm/Hz^{1/2}$; the corresponded pictures are given
on {\ref{fig4}}. The sensitivity was limited by the resonance thermal 
noise of the bar and approximately the same level of optical noises. This 
result is two orders of magnitude less the designed level 
$\sim 10^{-16}cm/Hz^{1/2}$.

%\begin{figure}[h]
%\centerline{\hbox{\includegraphics[width=0.4\textwidth]{rudenko_fig4a.eps}
%\hspace*{1cm}
%\includegraphics[width=0.4\textwidth]{rudenko_fig4b.eps}%
%}}
%\caption{Spectral characteristics of OGRAN pilot model:\protect\\
%a) frequency noise distribution, b) displacement sensitivity.}
%\label{fig4}
%\end{figure}

At present a new 3\,Wt single mode stabilized laser is under preparation 
as well 
as mirrors with high reflectivity $R=0.9997$ and small losses $\sim 50\,ppm$. 
This modification must allow to get the $1.5$ orders improvement in the 
registering amplitude for the pilot model out of the acoustical resonance.
Meanwhile big components of the real set up partly are ready (vacuum
chamber, bars) partly are in progress as well as an infrastucture of 
the project: the undeground lab in Baksan, factory hangar in SAI MSU for a
test measurements with big bars etc.

The plan of development foresees (first) an operation with the OGRAN set up 
at the level $\sim 3\cdot 10^{-19}$ with ``room temperature bar" 
starting from 2006 and (second) a parallel construction of the cryogenic 
OGRAN version with the final goal $h\sim 3\cdot 10^{-22}$. In this activity 
OGRAN collaboration hope to use a large experience and assistance of
the Italian cryogenic bar group (ROG collaboration \cite{ref2}).

\section*{Acknowledgements}
Authors appreciate a decisive role of Directors of Institutions mentioned
in the title, --- academicians Sergei Bagaev, Viktor Matveev and Anatolij 
Cherepashchuk and gratitude them for their inspitration and providing a
financial supportion of the Project. Many very usefull discussions were 
carried out with our colleagues professors V.~Braginskii, A.~Manukin, 
V.~Rubakov, V.~Lipunov, K.~Postnov, G.~Kogan.

This research was supported by two national programs (contracts of Ministry
of Sciences): Neutrino Detectors, Fundamental Metrology and partly by the
Grant of RFBR N 04-02-17320.

Our final note is sad: we should express a very deep feeling associated 
with the loss in a tragic accident our young, very talented colleague and 
co-author Andrej Serdobolskii who made a valuable contribution in the 
OGRAN prototype and was very good friend for us.

\section*{Figure captions}

\begin{itemize}

\item[{\large Figure 1.}] Principle layout of the OGRAN detector.

\item[{\large Figure 2.}] LPF-scheme of laser frequency stabilisation:

EOM --- electro-optical phase modulator;

DBM --- double balance mixer;

AFC --- laser frequency driver;

PD --- photodetector.

\item[{\large Figure 3.}] The principal layout of OGRAN project:

EOM --- electro-optical modulator;

45P --- polarizer;

$\lambda /4$ --- plate $\lambda/4$;

FD --- photodetector;

Mrs --- adjustable mirrors.

\item[{\large Figure 4.}] Spectral characteristics of OGRAN pilot model:

a) frequency noise distribution, b) displacement sensitivity.

\end{itemize}


\newpage

\begin{thebibliography}{10}
\bibitem{ref1} URL: {\it http://igec.lnl.infn.it }.
\bibitem{ref2} ROG, P.\,Astone et. al., {\it Astropart. Phys.}, {\bf 10}, 83 (1999). 
\bibitem{ref3} LIGO, {\it http://www.ligo.org }.
\bibitem{ref4} VIRGO, {\it http://www.virgo.infn.it }.
\bibitem{ref5} GEO600, B.\,Willke, et. al., {\it Class. Quant. Grav.}, {\bf 19}(7), 1377 (2002). 
\bibitem{ref6} TAMA300, H.\,Tagoshi et. al., {\it Phys. Rev. D}, { \bf 63}, 062001 (2001). 
\bibitem{ref7} GRAIL, G.\,Frossati, {\it J. Low Temp. Phys.}, {\bf 101}, 81 (1995); preprint gr-qc/9804073.
\bibitem{ref8} SPHERA, E.\,Coccia, J.A.\,Lobo, J.A.\,Ortega. {\it Phys. Rev.}, {\bf 52}, 3735 (1995).
\bibitem{ref9} O.D.\,Aguiar, L.A.\,Andrade, J.L.\,Barroso et. al., {\it Class. Quant. Grav.}, {\bf 21}, 457 (2004).
\bibitem{ref10} E.\,Amaldi, P.Bonifazi, M.G.Castellano et. al., {\it Europhys. Lett.}, {\bf 3}, 1325 (1987). 
\bibitem{ref11} P.\,Astone, D.\,Babusci, M.\,Bassan et. al., {\it Class. Quant. Grav.}, {\bf 19}, 5449 (2002). 
\bibitem{ref12} J.\,Weber, {\it Found. Phys.}, {\bf 14}, 1185 (1984). 
\bibitem{ref13} G.\,Preparata, {\it Il Nuovo Cimento}, {\bf B101}, 625 (1988). 
\bibitem{ref14} R.\,Sisto, A.\,Moleti, {\it Int. J. Mod. Phys.}, {\bf 13}, 4, 625 (2004). 
\bibitem{ref15} J.T.\,Whelan, E.\,Daw, I.S.\,Heng et. al., preprint gr-qc/0308045.
\bibitem{ref16} V.N.\,Rudenko, {\it Astronomy Reports}, {\bf 78}, 12, 1116 (2001). 
\bibitem{ref17} V.\,Kulagin, A.\,Polnarev, V.\,Rudenko, {\it JETP}, {\bf 64}, 6, 915 (1986). 
\bibitem{ref18} A.\,Gusev, V.\,Kulagin, V.\,Rudenko, {\it Gravitation and Cosmology}, {\bf 2}, 1, 68 (1996).
\bibitem{ref19} L.\,Conti, G.A.\,Prodi, S.\,Vitale et. al., {\it ``Gravitational waves and experimental gravity". Proc. 34th Ren.
Moriond}, World Publish, p. 85 (2000).
\bibitem{ref20} S.N.\,Bagaev et. al., {\it XI Conf. Laser Optics 2003}, S.-P.
30.06.04, Technical Program (Abstracts), Publ. OSA, FrPD-07 (2003).
\bibitem{ref21} J.\,Weber, {\it Phys. Rev.}, {\bf 117}, 306 (1960); the book ``General
relativity and gravitational waves". NY, 1962.
\bibitem{ref22} {\it http://www.ligo-wa.caltech.edu }.
\bibitem{ref23} {\it http://www.ligo-la.caltech.edu }.
\bibitem{ref24} B.\,Abbott, R.\,Abbott, R.\,Adhikari et. al., preprint gr-qc/0312056.
\bibitem{ref25} B.\,Abbott, R.\,Abbott, R.\,Adhikari et. al., preprint gr-qc/0308069.
\bibitem{ref26} V.M.\,Lipunov, preprint astro-ph/0406502.
\bibitem{ref27} B.\,Abbott, R.\,Abbott, R.\,Adhikari et. al., preprint gr-qc/0308050.
\bibitem{ref28} B.\,Abbott, R.\,Abbott, R.\,Adhikari et. al., preprint gr-qc/0312088. 
\bibitem{ref29} K.\,Compton, D.\,Nicolson, B.\,Schutz, {\it Proc. MG-7}, Stanford Univ. USA, W. Sci. Publ., 1078
(1994).
\bibitem{ref30} P.\,Astone, K.M.\,Borkowskii, P.\,Jaranowski, A.\,Krolak, {\it Phys. Rev. D}, {\bf 65}, 022001
(2002).
\bibitem{ref31} P.\,Astone et. al., {\it Phys. Rev. D}, {\bf 65}, 022001 (2002).
\bibitem{ref32} E.\,Mauceeli, M.P.\,McHugh, W.O.\,Hamilton et. al., preprint gr-qc/0007023.
\bibitem{refA3} P.\,Astone et. al., {\it Phys. Let. B}, {\bf 385}, 421 (1996)
\bibitem{ref3A} P.\,Astone et. al., {\it Astron. Astrophys.}, {\bf 351},  811 (1999). 
\bibitem{ref33} L.\,S.Finn, S.D.\,Mohanty, J.D.\,Romano,  {\it Phys. Rev. D}, {\bf 60}, 121101 (1999).
\bibitem{ref34} L.\,Grishchuk, V.\,Lipunov, K.\,Postnov et. al., {\it Physics-Uspehi}, {\bf 44}, 1 (2001).
\bibitem{ref35} T.\,Piran, {\it Phys. Rep.}, {\bf 314}, 575 (1999).
\bibitem{ref36} A.V.\,Gusev, V.K.\,Milyukov, V.N.\,Rudenko et. al., {\it ``Gravitational
waves'', Proc. 2th E. Amaldi Conf.}, World Sci., 512 (1998).
\bibitem{ref37} M.T.\,Murphy, J.K.\,Webb, I.S.\,Heng, {\it MNRAS}, {\bf 316}, 657 (2000). 
\bibitem{ref38} V.N.\,Rudenko, {\it ``Gravitational waves and experimental gravity", Proc. 34th Ren.
Moriond}, World Publish., 269 (2000).
\bibitem{ref39} G.\,Modestino, A.\,Moleti, {\it Phys. Rev. D}, {\bf 65}, 022005 (2002). 
\bibitem{ref40} P.\,Bonifazi, G.V.\,Pallottino, A.V.\,Gusev et. al., {\it Astron. Astroph. Trans.}, {\bf 22}, 4-5, 557 (2003).
\bibitem{ref41} P.\,Astone et. al., {\it Phys. Rev. D}, {\bf 66}, 102002 (2002).
\bibitem{ref42} P.\,Tricarico et. al., {\it Phys. Rev. D}, {\bf 63}, 082002 (2001).
\bibitem{ref43} P.\,Astone et. al., preprint astro-ph/0408544.
\bibitem{ref44} C.A.\,Dickson, B.F.\,Schutz, {\it Phys. Rev. D}, {\bf 51}, 6, 2644 (1995).
\bibitem{ref45} V.\,Rudenko, A.\,Gusev, V.\,Kravchuk, M.\,Vinogradov, {\it JETP}, {\bf 91}, 5, 845 (2000). 
\bibitem{ref46} P.\,Astone et. al., {\it Class. Quant. Grav.}, {\bf 19}, 5449 (2002). 
\bibitem{ref47} S.\,Finn, {\it Class. Quant. Grav.}, {\bf 20}, L37 (2003). 
\bibitem{ref48} P.\,Astone et. al., {\it Class. Quant. Grav.}, {\bf 20}, 785 (2003). 
\bibitem{ref49} E.\,Coccia, F.\,Dubath, M.\,Maggiore, preprint gr-qc/0405047.
\bibitem{ref50} LCGT: K.\,Kuroda et. al., {\it Int. J. Mod. Phys.}, {\bf 8}, 5, 557 (1999). 
\bibitem{ref51} P.\,Astone et. al., {\it Astroparticle Phys.}, {\bf 7}, 231 (1997). 

\end{thebibliography}
\end{document}